\documentstyle[12pt]{article}

\begin{document}
\pagestyle{plain}
\begin{center}
{\Large \bf Theoretical Physics Institute}\\
\end{center}
\begin{flushright}
{TPI-MINN-94/23-T}
\\
{UMN-TH-1261}\\
{Technion-PH-94-8}\\
{September 1994}\\
\end{flushright}
\vspace{.2cm}
\begin{center}

{\Large \bf Calculation of $1/m^3_c$ terms in the total \\
semileptonic width of D mesons.
}
\end{center}

\vspace{0.4 cm}
\begin{center}
{\large Boris Blok}

\vspace{0.2 cm}
{Department of Physics\\ Technion -- Israel Institute of
Technology\\
Haifa 32000, Israel
} \\

 \vspace{0.2 cm}
 and\\
 \vspace{0.2 cm}
{\large R. David Dikeman and M. Shifman}

\vspace{0.2 cm}
{ Theoretical Physics Institute,
 University of Minnesota \\ Minneapolis, MN 55455, USA}\\

 \end{center}
\vspace{0.4cm} \noindent

\begin{abstract}

\noindent

We calculate the 1/$m^3_c$ corrections in the inclusive semileptonic
widths of $D$ mesons. We show that these are due to the novel
penguin
type operators that appear at this level in the transition operator.
Taking into
account the nonperturbative
corrections leads to the predicted value of the semileptonic width
significantly lower than the experimental value. The $1/m^3_c$
worsen the situation or at the very least, within uncertainty,
give small contribution. We indicate possible
ways out. It seems most probable that violations of duality are
noticeable
in the energy range characteristic to the inclusive decays in the
charm family.
Theoretically these deviations are related to divergence of the
high-order
terms in the power expansion in the inverse heavy quark mass.

\end{abstract}

\thispagestyle{empty}
\endtitlepage
\newpage

\setcounter{page}{1}

\section{ Introduction}

 Recently a QCD based approach to calculation of total inclusive
decay rates of heavy ($B$ and $D$) mesons was developed \cite{1} --
\cite{6}.
The approach is based
   on the systematic expansion in the inverse heavy quark mass
within  the
operator product expansion (OPE) \cite{7}. In this note we will
discuss
application of this formalism to the calculation of the total
 semileptonic inclusive width of $D^0$ mesons. Unlike the previous
model
calculations the OPE-based method
 gives us full control over all relevant parameters in theoretical
 expressions.

 The leading perturbative
  (${\cal O} (\alpha_s)$) \cite{8,9} and nonperturbative (${\cal
O}(1/m^2_c))$
\cite{4}
corrections  have been found previously.
 It turns out that under a reasonable choice of the $c$-quark mass
the
 predicted value of $\Gamma_{\rm sl}(D)$ is significantly lower than
the
 corresponding experimental number \cite{10}. We calculate the
next-to-leading  nonperturbative correction of order $1/m_c^3$ and
show that
it only worsens the situation, or at the very least, gives small
contribution
within uncertainty limits.
 We then indicate possible ways out.

At the level of ${\cal O}(1/m^3_c)$ terms there
arise penguin diagrams generating new, four-fermion operators of
dimension
6. The penguin graphs were introduced 20 years
ago \cite{penguin} in the strange particle decays where it was crucial
that
they produce right-handed quarks. In the $D$ meson semileptonic
decays the
origin of penguins is quite different -- they appear  at the
level of the {\em transition operator} and   give rise
to contribution of the annihilation type. Usually it is believed
that the latter is suppressed by chirality arguments. The suppression
is
lifted, however, due to the fact that penguins produce the
right-handed
quarks, much in the same way as in Ref. \cite{penguin}.

Since the $1/m_c^3$ terms do not eliminate the discrepancy between
the
theoretical prediction for $\Gamma_{\rm sl} (D)$ and experiment
a natural question immediately comes to one's mind: what went
wrong?
In estimating the $D$-meson matrix elements of the four-fermion
operators
we use factorization. In the limit of large number of colors,
$N_c\rightarrow\infty$, this approximation becomes exact. One may
suspect,
however, that at $N_c=3$ deviations from factorization are
substantial.
Can these deviations be a solution of the problem?

Although logically this possibility is not ruled out {\em a priori}
it is hard to believe that this is the case. Indeed, if the problem is to
be solved
in this way not only the matrix element of dimension-6 operators
must be
enhanced by a factor of $\sim 3$, its sign has to be reversed as
compared
to what one obtains within factorization.

The second logical possibility - an enhanced contribution coming
from
dimension-7 operators - also seems very unlikely.

Thus, we are inclined to conclude that the failure of the standard
$m_Q^{-1}$
expansion in the case of $\Gamma_{\rm sl} (D)$ is due to the fact
that the
charmed quark mass is too light for duality to set in. This assertion
will be
explained in more detail in Sect. 4. Here we only note that duality is
one of
the crucial elements of the calculation of the inclusive widths within
the
heavy
quark expansion. Theoretically the onset of duality is related to the
behavior
of high-order terms in the $1/m_Q$ expansion, the divergence of the
$1/m_Q$ series. The fact that the OPE-based power expansions are
actually asymptotic is well established \cite{Shifman}. Very
little is
known, however, about specific details of the divergence.

An indirect although a very strong argument that the charmed quark
is only at
the border, or even below the boundary, of the duality domain comes
from
consideration of the lifetime hierarchy in the charmed family (for a
recent
discussion see \cite{Blok}). Although ${\cal O}(m_c^{-2})$ and
${\cal O}(m_c^{-3})$ terms qualitatively reproduce the observed
pattern
some of the predicted lifetime ratios (which span an order of
magnitude!)
are off by a factor $\sim 2$. The predicted ${\cal O}(m_c^{-2}, m_c^{-
3})$
deviations from the asymptotic limit are typically smaller than what
is
observed experimentally. Needless to say that asymptotically all
lifetimes are
equal.

As was  mentioned above, the issue of the inclusive semileptonic $D$
decays
 was addressed in the recent literature more than once. The approach
to the
problem accepted  e.g. in Ref.
\cite{10} is inverted. It is assumed that the theoretical
 prediction for $\Gamma_{\rm sl} (D)$ truncated at the leading order
of
perturbation theory and at the leading order of the $1/m_c$
expansion
(i.e. keeping only $1/m_c^2$) is accurate enough to use it to fit the
values of
the quark mass and other theoretical parameters from
$\Gamma_{\rm sl}
(D)_{\rm exp}$. The value of the charmed quark mass emerging in
this way is
unrealistic. At the same time
 the
average value of the heavy quark kinetic energy $\mu^2_\pi$
remains
essentially
undetermined. We, instead,  use the best available scientific
estimates  of
$m_c$ and $\mu^2_\pi$. We will see that the results are
in direct disagreement with the experimental data.

Organization of the paper is as follows.
In section 2, we discuss the current situation. The naive parton result
is
augmented
by its perturbative - to $\alpha_s$ -  and non-perturbative - to
$1/m^2_c$ -
corrections. Section 3 shows the situation with $1/m^3_c$ non-
perturbative
corrections. We will see that they don't improve the match with
experiment.
Finally, we discuss ways out of the dilemma.

\section{The starting point}

The theoretical expression for the inclusive semileptonic width of the
$D$
meson ($c\rightarrow s l\nu $  transition), including the leading
perturbative and non-perturbative corrections, has the form
\begin{equation}
\Gamma (D\rightarrow l\nu X_s)=\frac{G^2_Fm^5_c}{192\pi^3}
|V_{cs}|^2
\left[ 1+ A^{(1)}\alpha_s - \frac{3\mu^2_G}{2m_c^2}-
\frac{\mu^2_\pi}{2m_c^2}\right] \, .
\label{width}
\end{equation}
Here $m_c$ is the charmed quark mass, and we neglected the
strange quark mass; $V_{cs}$ is the
corresponding CKM matrix element. The coefficient $A^{(1)}$ of the
${\cal
O}(\alpha_s )$ term
has been known for many years, see Ref.
\cite{Cabibbo} whose authors merely adapted the QED radiative
correction to $\mu\rightarrow e\nu\bar\nu$ (the
original
QED calculations are published in Ref. \cite{QED}).
The explicit expression for $A^{(1)}$ depends on what definition of
the
quark mass $m_c$ is accepted. The straightforward borrowing from
$\mu\rightarrow e\nu\bar\nu$ implies the use of the so called pole
mass. Although this parameter is not well-defined in full QCD (see
Ref. \cite{pole} and the  discussion below) it is admissible for a
limited  technical purpose of presenting the ${\cal O}(\alpha_s)
$ correction. Then
\begin{equation}
A^{(1)} = -\frac{2}{3\pi}(\pi^2-\frac{25}{4})\, .
\label{pert}
 \end{equation}

The leading non-perturbative correction in the $1/m_c$ expansion
for $\Gamma (D\rightarrow l\nu X_s)$ was first calculated in Ref.
\cite{4}. The term of the first order in $1/m_c$ is absent, the so
called
CGG/BUV theorem \cite{3,4}. At the level ${\cal O}(m_c^{-2})$ the
correction is determined by two parameters,
$$
\mu_G^2
=\frac{\langle D|(i/2)\sigma_{\mu\nu}G^{\mu\nu}|D\rangle}{2M_D}
\approx \frac{3}{4}(M^2_{D^*}-M^2_D)\, ;
$$
\begin{equation}
\mu_\pi^2
=\frac{\langle D|(i\vec D)^2|D\rangle}{2M_D} \, .
\label{mugmupi}
\end{equation}

Before proceeding to numerical estimates it is worth discussing the
parameters in Eq. (\ref{width}) in more detail. Numerically the most
important parameter is the quark mass since it enters in the fifth
power.
As was shown in Refs. \cite{3} -- \cite{6} it is the current quark
masses,
not the constituent one and not the hadron mass, that appear in the
systematic $1/m_Q$ expansions. If we limit ourselves to the
${\cal O}(\alpha_s ) $ expression for $\Gamma $ and do not ask any
questions
about higher-order terms we are free, of course, to express the result
in terms of the pole mass or in terms of the running mass normalized
at any point we like -- this will merely redefine the coefficient in
front
of $\alpha_s$ in a certain  way. We are aimed, however, at better
accuracy; in particular, we want to include the power
non-perturbative terms
in analysis.

The use of the pole mass is generically inconsistent; only
the running mass can appear in any OPE-based expression, see
\cite{pole}.
The question is what normalization point is relevant. The full
expression
for the decay rate, including all terms in the $\alpha_s$ expansion, is
certainly
independent
of the choice of the normalization point $\mu$, an auxiliary
parameter
in the operator product expansion. For a truncated series --
and we are forced, of course, to truncate the series at the level of the
leading,
or, at best, the next-to-leading term -- the choice of $\mu$ is
important since under a ``natural" choice the coefficients
in the neglected part of the series are small while under  ``unnatural"
choices
they can be abnormally large. In Ref. \cite{pole} it was explicitly
shown that
the natural choice for $m_c$ is $\mu =\mbox{Const.}m_c$ (see also
\cite{BBZ,N}).
The leading operator in the expansion in the
problem at hand is $\bar c (i\not\!\! D)^5 c$. By adopting the
normalization
point $\mu =\mbox{Const.}m_c$ we avoid any large corrections.
Non-perturbative effects enter through the matrix element of this
operator; they are also represented by matrix elements of other
(subleading) operators, for instance, $\bar c (i\not\!\! D)^3i\sigma G
c$.

Using the
equations of motion one reduces the leading  operator to $m_c^5\bar
c c $,
where both $m_c$ and $\bar c c$ are taken at  $\mu = m_c$. We
then evolve $\bar c c $ down to a low normalization point,
$\mu\ll m_c$; the net effect of this evolution is reflected in a factor
of the type $c(\mu, m_c) =1+ a_1(m_c/\mu )\alpha_s (m_c) + a_2
(m_c/\mu )\alpha_s^2(m_c) + .... $
which is, anyway, included in the perturbative calculation, see Eq.
(\ref{width}) \footnote{In principle, this
factor must contain also  terms of order $(\mu /m_c)^n$
due to the exclusion of the domain below $\mu$ from the
perturbative calculation. It is important that the power $n$ starts
from $n=2$, and  this term conspires with
$\mu_G^2$ and $\mu_\pi^2$.}.  Once the operator $\bar c c$ is
evolved
down to a low normalization point we use\footnote{The O($m_c^{-
2}$) terms in
Eq.(4) were derived in Ref. 4. The fact that the operators of
dimension 6
are absent in this expansion can be easily established by using the
equation
of motion, $$\frac{1-\gamma_0}{2}c = \frac{1}{2m_c}\hat{\pi}c$$
Then $$\bar{c}\frac{1-\gamma_0}{2}c=\bar{c}\frac{1-
\gamma_0}{2}\frac{1-
\gamma_0}{2}c = \frac{1}{4m_c^2}\bar{c}\hat{\pi}\hat{\pi}c$$
implies that
$$\bar{c}(1-\gamma_0)c = \frac{1}{2m_c^2}\bar{c}(\pi^2+\frac{i}{2}
\sigma G)c.$$
Moreover, using the equations of motion, again we see that $\bar{c}
\pi_0^2c$
actually reduces to an operator of dimension 7 and we arrive at Eq.
(4).}
the relation \cite{4}
\begin{equation}
\bar c c = \bar c \gamma_0 c + \frac{1}{4m_c^2}\bar{c}
i\sigma G c -
\frac{1}{2m_c^2}\bar{c}\vec{\pi}^2 c  +
{\cal O}(1/m_c^4)+\mbox{total derivatives}.
\label{cc}
\end{equation}

The numerical value of the (one loop) pole mass of the charmed
quark was
determined
long ago from the charmonium sum rules \cite{Novikov},
$m_c^{pole}\approx 1.35$ GeV (see also \cite{SVZ}). A recent advent
of HQET
\cite{HQET} allows one to conduct a consistency check of this
estimate.
Indeed, let us observe that
\begin{equation}
 m_b-m_c\,= \, \bar M_B - \bar M_D
\,+ \,
\mu_\pi^2  (\frac{1}{2m_c}-\frac{1}{2m_b}) \,+\, {\cal
O}(1/m_c^3\,,\;
1/m_b^3) \;\;.
\label{mdif}
\end{equation}
where
$$
\bar M_B =
\frac{M_{B}+3M_{B^*}}{4}
$$
and the same for $D$. Next,
for the pole mass of the $b$ quark a very precise evaluation
\begin{equation}
m_b = 4.83\pm 0.03 \,\, {\rm GeV}
\label{mb}
\end{equation}
is obtained in the recent analysis of
the QCD sum rules
for the $\Upsilon$ system \cite{voloshin2}. To be on a safe side we
multiplied the original error bars by a factor of 4.  It is worth
noting that it is very difficult -- practically impossible --
to go outside the indicated limits. The central value of
$m_b$ above implies $m_c \approx 1.33$ GeV (provided
we accept the estimate of Ref. \cite{Ball} for $\mu_\pi^2$, see below).
The most generous error bars in $m_b$ and $\mu_\pi^2$ (see below)
Plus or
are translated in $\pm 70$ MeV uncertainty in $m_c$. It seems
perfectly
safe to say that $m_c^{pole}$ lies between 1.25 and 1.40 GeV --
one can not imagine that the one-loop pole $c$ quark mass
is less than $1.25$ or larger than $1.40$.
The HQET result $m_c^{pole}\approx 1.33$ GeV matches very well
the QCD sum rule number quoted above.

As was mentioned, in the purely theoretical aspect, it is more
consistent to use the
running mass in the
expression for the total semileptonic decay rate. One may choose
the so called Euclidean mass (\cite{Novikov,SVZ}) or the running
$\overline{MS}$ mass evaluated at $p^2= -m_c^2$.  Both are close to
each other
numerically, and are smaller numerically than $m_c^{pole}$ since
they
are deprived of a part of the (perturbative) gluon cloud compared to
$m_c^{pole}$,
\begin{equation}
m_c^{pole} = m_c^{eucl}\left[ 1+\frac{2\ln 2\alpha_s }{\pi } +
...\right]\quad \mbox{and}\quad
 m_c^{pole} = m_c^{\overline{MS}}(m_c)\left[ 1+\frac{4\alpha_s}{3\pi
} +
...\right]\, .
\label{mass}
\end{equation}
For instance, a fit in the charmonium sum rules yields
\cite{Novikov} $m_c^{eucl}\approx 1.25$ GeV.
 Since our task is limited -- in the numerical evaluation we will
not go beyond the first order in $\alpha_s$ -- essentially it does not
matter
which expression for the width is used: the one written in terms of
the pole
mass or in terms of any other mass from Eq. (\ref{mass}).

Let us first ignore the correction terms in Eq. (\ref{width})
altogether. Then the naive parton-model expression
$$
\Gamma_0 (D\rightarrow l\nu X_s)=\frac{G^2_Fm^5_c}{192\pi^3}
|V_{cs}|^2
$$
with $m_c=1.4$ GeV yields $\Gamma (D\rightarrow l\nu X_s)=
1.1\times 10^{-13}$ GeV, to be compared with
\begin{equation}
\Gamma (D\rightarrow l\nu X_s)_{\rm exp}
=1.06\times 10^{-13}\,\,\, \mbox{GeV}.
\label{exp}
\end{equation}
The value of $m_c=1.4$ GeV is at the upper boundary of what we
believe is
allowed for the pole charmed quark mass.  We will consistently push
the estimates of $\Gamma (D\rightarrow l\nu X_s)$ to the high side
and,
hence, use this value for orientation.

If, instead, $m_c=1.25$ GeV is substituted then the parton-model
formula
gives 0.6 of $\Gamma (D\rightarrow l\nu X_s)_{\rm exp}$.

The key point, emphasized in the introduction, is that all known
corrections, perturbative and non-perturbative are negative.
Consider
first the ${\cal O}(\alpha_s )$ correction in Eq. (\ref{width}). To
evaluate it
numerically one needs to know the normalization point of
$\alpha_s$.
Usually it is taken to be $\mu\sim m_c$. Again this is the question of
higher-order terms. Recently it was argued \cite{9} within the
Brodsky-Lepage-Mackenzie hypothesis \cite{BLM} that the ${\cal O}
(\alpha_s^2)  $ terms are negative and large, so that actually the
normalization point of $\alpha_s$ constitutes a relatively small
fraction of
$m_c$. Since this effect works in the direction of reducing
$\Gamma (D\rightarrow l\nu X_s)_{\rm theor}$ and we agreed to
push
the estimate to the high side, we will ignore the
${\cal O}
(\alpha_s^2)  $ terms
remembering, however, that our estimate will then lie higher than
the actual theoretical prediction.

For consistency we use the
value $\Lambda_{\rm QCD}$
 $\sim 150$ MeV relevant  to the one-loop approximation.
Then $\alpha_s (m_c) \approx 0.30$, which is consistent with the
recent very precise evaluation \cite{voloshin2}, as well as with the
previous analyses of data on deep inelastic scattering \cite{Alta}.
(The  corresponding two-loop value $\Lambda_{\rm QCD}\sim 250 $
MeV). Notice that bigger
values of $\Lambda_{\rm QCD}$ sometimes advocated in the
literature only worsen
 the disagreement between theory and experiment. With this value
of $\alpha_s (m_c)$ we find that the ${\cal O}(\alpha_s )$ correction
in Eq. (\ref{width}) is equal to $-0.24 $, i.e. it further reduces
$\Gamma (D\rightarrow l\nu
X_s)_{\rm theor}$ by a quarter!

Let us discuss now the non-perturbative ${\cal O}(m_c^{-2}) $ terms.
The value of $\mu_G^2$ is known phenomenologically, see Eq.
(\ref{mugmupi}),
\begin {equation}
\mu_G^2 \approx 0.41 \,\, \mbox{GeV}^2\, ,
\label{mug}
\end{equation}
 (small effects due to the anomalous dimension of the
chromomagnetic
operator \cite{Falk}
are neglected). As for $\mu_\pi^2$, at present this parameter is  not
measured,
although it is measurable in principle \cite{optical}. At least two
independent
lower bounds are established \cite{voloshin3,optical,third} which
turn out to
be close numerically,
$$
\mu_\pi^2 >0.4\, \mbox{GeV}^2 \, .
$$
Moreover, the value of $\mu_\pi^2$ was evaluated in the QCD sum
rules
\cite{Ball},
with the result
\begin{equation}
\mu_\pi^2 = 0.6 \pm 0.1\, \mbox{GeV}^2 \, .
\label{mupi}
\end{equation}
Now we are finally able to estimate the ${\cal O}(m_c^{-2})$ terms in
$\Gamma (D\rightarrow l\nu X_s)_{\rm theor}$. Again, trying to
increase
$\Gamma_{theor}$ we use the lower value of
$\mu_\pi^2$;
we then conclude that the chromomagnetic and kinetic term
contribute to the brackets in Eq. (\ref{width}) $-0.31$ and
$-0.13$, respectively.
So collecting everything, we have (to
${\cal O}(\alpha_s )$ and  ${\cal O}(m_c^{-2} )$)
$$\Gamma (D\rightarrow l\nu X_s)_{\rm theor}= \Gamma_0 [1 -
0.24 - 0.31 - 0.13]
.$$
We have less than
half of the
experimental
width! Can the ${\cal O}(m_c^{-3} )$ terms fix the situation?
As we will see in the next section the answer to this question is
negative.

\section{The $1/m^3_c$ corrections to the semileptonic width}

Since the $1/m^2_c$ terms do not solve the problem of the total
semileptonic width, it is natural to consider the corrections due to
$1/m^3_c$ terms.  They can be calculated in the standard way
within the heavy quark expansion \cite{1} -- \cite{6}. Below
the basic points of derivation are sketched.

We start  from
the  weak Lagrangian describing the semileptonic decays
\begin{equation}
{\cal L}(\mu) = \frac{G_F}{\sqrt{2}}V_{sc}{\cal O}\, .
\label{9}
\end{equation}
Here
\begin{equation}
{\cal O} = (\bar s\Gamma_\mu c)(\bar \nu\Gamma^\mu e)  \,\,  ,
\label{10}
\end{equation}
and $\Gamma_\mu = \gamma_\mu (1+\gamma_5)$.  Eqs. (\ref{9}),
(\ref{10}) present
the lagrangian relevant to the $c\rightarrow s\bar e \nu$ transition.

\par Next, we construct the transition operator
$\hat T(c\rightarrow X\rightarrow c)$,
\begin{equation}
\hat T = i\int d^4x T\{ {\cal L}(x) {\cal L}(0)\} =
\sum_i C_i {\cal O}_i
\label{eq:expansion}
\end{equation}
describing the diagonal amplitude
with the heavy quark $c$ in the initial and final states (with identical
momenta).
 The transition operator $\hat T$ is built by means of OPE
as
an expansion in local operators ${\cal O}_i$.
The lowest-dimension operator in
$T(c\rightarrow X\rightarrow c)$ is $\bar c c$, and the complete
perturbative prediction corresponds to the perturbative calculation
of the
coefficient of this operator.  In calculating the coefficient of $\bar c
c$ we treat the light  quarks in $X$ as hard and neglect the
soft modes. Say, we ignore the fact that in a part of the phase space
the $s$ quark line is soft and can not be treated perturbatively.
Likewise, we ignore interaction with the soft gluons.
The presence of the soft quark-gluon ``medium" is reflected
in higher-dimensional operators.

Once the expansion (\ref{eq:expansion}) is built we average $\hat T$
over the
$D$ meson to obtain the lifetime,
$$\Gamma = \frac{1}{M_{D}}{\rm Im}<D|\hat{T}|D>.$$   At this stage
the
non-perturbative
large-distance dynamics enter through the matrix elements of the
operators of dimension  5 and higher. There are
no operators of dimension 4 \cite{3}. The operators of dimension 5
has been already discussed. The only new operators relevant at the
level of dimension 6 are the four-fermion operators of the
type
$$
{\cal O}_6 =\bar{c} \Gamma q \bar{q} \Gamma c
$$
where $q$ generically denotes the light quark field and $\Gamma$
stands for a  combination of $\gamma$ and $t_a$ matrices. What
particular
combination
is
relevant will be seen from what follows. There are two distinct
sources of
$1/m_c^3$  corrections in the total semileptonic width: operators of
dimension six arising in the expansion for $\hat{T}$, and $1/m_c$
corrections in the $D$-meson matrix elements of the operator
$\bar{c}\sigma Gc$
and $\bar{c}c$. Let us discuss these terms starting from dimension 6
operators
in the expansion for $\hat{T}$. We shall see that since $\Gamma$ is a
Lorentz
scalar quantity, only Lorentz scalar operators can contribute. Thus,
the only
operator showing up at this level will be of the four-fermion type,
${\cal O}_6.$

\subsection{The four-fermion operators at order $\alpha_s^0$ and in
LLA}

A four-fermion operator appears in $\hat T$ in the zeroeth order in
$\alpha_s$
from the diagram of Fig. 1.
The corresponding result can be read off from Eq. (17c) of Ref.
\cite{VS}, where one must put $C_+=C_- =1$ and eliminate the
color factor of 3 from the numerator,
\begin{equation}
{\rm Im} \  {\hat T}^{(0)}
=-\frac{G_F^2m_c^2}{8\pi}|V_{cs}|^2
\left\{ ({\bar c}_i\Gamma_\mu c_k -(2/3) {\bar c}_i\gamma_\mu
\gamma_5 c_k)({\bar s}_k\Gamma_\mu s_i)\right\} \, .
\label{Tzero}
\end{equation}
The expression in the braces includes the left-handed $s$ quarks
only. If we use the standard factorization procedure (i.e.
saturation by the vacuum state) for estimating the matrix
element of this operator over the $D$-meson state we get zero for
two reasons. First, in the factorization approximation
${\hat T}^{(0)}$ corresponds to the annihilation contribution,
$c\bar s \rightarrow l\nu$, which shows up only in $D_s$.
Second, even for $D_s$ the chiral structure is ``wrong", and after
factorization
the matrix element $\langle D|{\hat T}^{(0)} |D\rangle$ vanishes.

Therefore, the four-fermion operator appearing in Eq. (\ref{Tzero})
by itself is not interesting. Let us recall, however, that
the normalization point of this operator in Eq. (\ref{Tzero})
is $\mu = m_c$, and before estimating its matrix elements we must
evolve it down to a low normalization point. (Of course, it is desirable
to go to $\mu$ of order of the typical off-shellness of quarks inside
mesons. Clearly we can not do this since then  the perturbative
calculation of the coefficient functions becomes meaningless.
We will make a compromise and evolve down to $\mu\sim 0.5$ GeV
assuming that, on one hand, the coefficient functions are still
calculable and, on the other hand, the factorization procedure can be
used for obtaining the $D$-meson matrix element.)
It is straightforward to take this evolution into account in the
leading logarithmic approximation (LLA); as a matter of fact, we just
parallel the standard penguin analysis (see Fig. 2a). What is crucial
is that
this evolution brings in new four-fermion operators, of a different
flavor and chiral structure, whose matrix elements over $D^0$
and $D^+$ do not vanish within factorization.
Calculating the diagram of Fig. 2a with the logarithmic accuracy
we get \footnote{This result can also be extracted from Eq. (20)
of Ref. \cite{VS}.}
\begin{equation}
{\rm Im} \ \hat T^{(1)}
=-\frac{G_F^2m_c^2}{8\pi}|V_{cs}|^2
\left\{
-\frac{\alpha_s}{3\pi}\ln\frac{m_c^2}{\mu^2}
\left( \frac{2}{3}\bar c\Gamma_\mu t^a c
+\frac{1}{3}\bar c \tilde\Gamma_\mu t^a c\right)
\sum_q \bar q \gamma_\mu t^a q\right\} \, ,
\label{LLA}
\end{equation}
where
$$
\tilde\Gamma_\mu = \gamma_\mu (1-\gamma_5)\,
$$
and $t^a$ are generators of $SU(3)$,
$$
t^a =\frac{1}{2}\lambda^a\, .
$$
(Here $\lambda^a$ are the Gell-Mann matrices.) Notice that the light-
quark
current, $\sum\bar q \gamma_\mu t^a q$ is actually
$\frac{1}{g}D_\nu
G_{\mu\nu}$; it includes all light-quark flavors and both,
left-handed and right-handed fields. This is a typical feature of the
penguin contribution \cite{penguin}, leading to a non-vanishing
contribution of Eq. (\ref{LLA}) to the $D$-meson matrix elements
within factorization. As a matter of fact, we can omit
the left-handed part of the light-quark current,
since, as was explained above, after factorization the term with the
left-handed part of the current will vanish, i.e.,

\begin{equation}
\sum_q \bar q \gamma_\mu t^a q=
\sum_q\frac{1}{2}\left\{
\bar q \Gamma_\mu t^a q
+ \bar q \tilde\Gamma_\mu t^a q\right\}
\rightarrow
\frac{1}{2}\sum_q\, \bar q \tilde\Gamma_\mu t^a q\, .
\label{current}
\end{equation}

\subsection{Full ${\cal O}(\alpha_s )$ calculation}

Unfortunately, $\ln m_c^2/\mu^2$ is not a very large numerical
parameter and, hence, neglecting non-logarithmic terms may seem
unjustified. Therefore, instead of summing up all logarithms in LLA
(which can be readily done, though), it seems reasonable
to limit oneself to the ${\cal O}(\alpha_s )$ calculation including both
the logarithmic and non-logarithmic terms, the more so that
we need the result only for the purpose of orientation. We want to
convince ourselves that the contribution of dimension-6 operators to
$\Gamma (D\rightarrow X_s l\nu)$ is negative.

It is most convenient to carry out the full ${\cal O}(\alpha_s )$
calculation using the background field technique. There are two
versions of this technique -- the first one was exploited in the
context of the inclusive semileptonic decays e.g. in Ref.
\cite{Koyrakh}, the second version, based on the Fock-Schwinger
gauge, is reviewed in Ref. \cite{NSVZ}. Both versions can be applied
in the case at hand. Here, the latter one is more suitable.

The dimension 6 operators come from two sources.
To see this, we write
explicitly the expression for the
transition operator.
$$
\hat {T} = - \frac {G^2_F}{2}\int d^4x e^{-iqx} \bar{Q}(x)
\Gamma_{\mu}
S_q(x,0) \Gamma_{\nu} L_{\mu \nu}Q(0)
$$
Where $L_{\mu \nu}$ is the lepton loop, and $S_q$ stands for the
light
quark Green function.

As mentioned, in the expression we have the propagator for the $s$
quark. When
expanded in the Fock-Schwinger gauge this propagator is
\begin{eqnarray}
S_q (x,0) &=&
        \frac{-im^2}{4\pi^2}\frac{K_1(m\sqrt{-x^2})}{\sqrt{-x^2}}
        -\frac{m^2\hat{x}}{4\pi^2 x^2}K_2(m\sqrt{-x^2})    \nonumber\\
& &             +\frac{\tilde{G_{\rho\lambda}}}{8\pi^2}
           \frac{mK_1(m\sqrt{-x^2})}{\sqrt{-x^2}}
           (x_{\rho}\gamma_{\lambda}\gamma_5)+
    \frac{G_{\rho\lambda}}{16\pi^2} mK_0(m\sqrt{-
x^2})\sigma_{\rho\lambda}
                                                                  \nonumber \\
& &      + \frac{2}{3}g\frac{i}{32\pi^2}(2K_0(m\sqrt{(-x^2m)})
          D^\alpha G_{\alpha\beta}\gamma_\beta
       -(D^\alpha G_{\alpha\beta}
        \hat x x^\beta +                                          \nonumber\\
& &     x^\gamma x^\alpha D_\gamma
        G_{\alpha\beta}\gamma^\beta
        -3ix^\gamma x^\alpha D_\gamma \tilde
        G_{\alpha\beta}\gamma_\beta\gamma_5)
        \frac{mK_1(m\sqrt{(-x^2)})}{\sqrt{(-x^2)}}\big)+....
\label{eq:A1}
\end{eqnarray}
where {\it m} is the {\it s} quark mass (needed for infrared
regularization).
We will see that the first of the $1/m_c^3$ corrections will come
from inserting the second (free) term of Eq. (\ref{eq:A1})
into $\hat T$. The second correction
comes from terms of order $\it{DG}$ in the propagator. (Note that the
term
proportional to $\tilde{G}$ and $D\tilde{G}$ in the $\it{s}$-quark
propagator
yields
zero due to their Lorentz
structure.)

The third source for corrections, the matrix elements of
dimension 3 and 5, will be discussed later.

We start with the expression for $\hat T$ with the free term of the
$\it{s}$-quark
inserted. This term can be read off from the diagram of Fig. 3.
Here we use the expression with $\it{s}$-quark mass put to zero
from
the very beginning since the result is non-logarithmic (infrared-
stable).
Doing the arithmetic, it is easy to show that, in the Fock-Schwinger
gauge
$${\rm Im } \
\hat{T_0} = \frac {G_F^2 |V_{\rm cs}|^2 \bar {Q(0)}
p^4 \hat{p} Q(0)}{384\pi^3}$$
where
$$p_{\mu} = iD_{\mu} -gA_{\mu},$$
and,

$$ A_{\mu} = \frac{1}{2\cdot 0!} {x_\rho}{G_{\rho\mu}} +
\frac{1}{3\cdot 1!}
{x_\rho}{x_\alpha}({D_\alpha}{G_{\rho\mu}})+....$$

Here we are interested in $1/m_c^3$ corrections and thus look only
at the term proportional to $\it{DG}$ in $A_\mu$ .
Our strategy is to pull $A_\mu$ to the left since $A(0)=0$. When
doing this,
we create commutators which are easily evaluated using the explicit
expression for  $A_\mu$. Performing this procedure, we obtain
the following contribution to the transition
operator:
  \begin{equation}
\hat T|_{free}=-iG_F^2 |V_{\rm cs}|^2\frac{\alpha_s}{
48\pi^2}m^2_c\bar c
\gamma_\beta
t^ac\bar q\gamma_\beta t^aq
\label{eq:add}
  \end{equation}
where the subscript `free' means the free term of the
$\it{s}$-quark propagator.
Note, here we have used the equation of motion

   \begin{equation}
  D^\alpha G^a_{\alpha \mu}=-g\bar q\gamma_\mu t^aq.
  \label{eq:motion}
  \end{equation}
After factorization (see Section 3.4) this yields
$$
\frac {\Delta \Gamma}{\Gamma_0}= \frac {8\alpha_s \pi f_D^2
M_D}{9 m_c^3}
$$
where $f_D$ is the axial constant of the D meson.
This key constant, $f_D$, is not measured accurately enough so far,
although
some experimental results do exist. It seems reasonable to rely on
theoretical calculations which were done both on the lattice and in
QCD sum
rules (see Refs. [32] and [33], respectively), thus we choose
$f_D=170$ MeV so as to push our estimate
for $\Gamma_{sl}(D)$ to the high side. We also take
$\alpha_s =0.31$, and $m_c = 1.4$
GeV . Plugging in these numericals the above expression gives,
$$
\frac {\Delta \Gamma}{\Gamma} = 0.016.
$$

Next, we consider the diagram of Fig. 2b. Its contribution
to the transition operator is calculated in the Appendix.
Fig. 2b singles out the $\it{DG}$ terms in the background
field expansion of the quark Green's function, see Eq. (30).
The infrared cut-off in the logarithm is achieved by
ascribing a mass of $\mu$ to the  $s$ quark line.
The lepton part of the diagram, which is trivial, must also be
inserted, of course. After the Fourier transformation we get
\\
\\
$$
\hat T|_{\rm {DG}}
  =iG^2_F| V_{\rm cs}|^2\left\{
\frac{\alpha_s}{ 72\pi^2}m^2_c({\ln}\,
\frac{m^2_c}{\mu^2}+\frac{2}{3})\right\}\times
$$
\begin{equation}
 (2\bar c \Gamma_\beta t^ac+
 \bar c \tilde\Gamma_\beta t^a c)\sum
 \bar q\gamma_\beta t^a q\, .
 \label{eq:eff1}
  \end{equation}
Note that the coefficient in front of the logarithm matches
the one in Eq. (15), as it should, which was obtained through the
logarithmic
mixing.
Again, after factorization, (See section 3.4), we get
$$
\frac {\Delta \Gamma}{\Gamma} = -\frac {16 \pi \alpha_s}{9m_c^3}
({\ln}\,
\frac{m^2_c}{\mu^2} + \frac{2}{3}) f_D^2 M_D.
$$
which is, numerically,
$$
\frac {\Delta \Gamma}{\Gamma_0} = -0.08
$$
where we have ascribed a value of $0.5$ GeV to $\mu$, and used the
same values
for
the other parameters as above.
\subsection{$O(\frac{1}{m_c^3})$ terms from matrix
elements of dimension 3 and 5 operators}

As was mentioned above, the matrix elements of operators of
dimension 3
and 5 also give rise to the $1/m_c^3$ corrections in the total
semileptonic width. Let us first consider Eq. (4). The matrix element
of the
first term is exactly unity. The matrix elements of the second and
third terms
do contain $1/m_c$ corrections, which we discuss here.

The point is that Eq. (3) expressing $<D|\bar c\frac{i}{2} \sigma Gc|D>$
 in terms
of the D* D mass splitting is valid only to the leading (zero) order
in $1/m_c$. Let us observe that the spin splitting yielding
$M_{D^\ast}^2 - M_D^2$ is determined by the following terms in the
heavy quark
Hamiltonian:
\begin {equation}
\Delta {\cal H} = \frac{1}{2m_c} \vec{\sigma} \vec{B} +
\frac{1}{4m_c^2}
\vec{\sigma} \vec{E} \times \vec{\pi}
\label {eq:HHamil}
\end {equation}
where $\vec{B}$ and $\vec{E}$ are the chromoelectric and
chromomagnetic
fields, respectively, $\vec{B} = g \vec{B}^a t^a$ and $\vec{E} = g
\vec{E}^a
t^a.$
To the leading order,
$$\Delta_D \equiv \frac{3}{4}(M_{D^\ast}^2 - M_D^2) = -<\vec{\sigma}
\vec {B}> = 0.405 \ {\rm { GeV}}^2$$
where $<\cdot \cdot \cdot>$ by definition means $(2M_D)^{-
1}<D|\bar{c}
\cdot \cdot \cdot c|D>.$
At the level of $1/m_c$ the second term in the heavy quark
hamiltonian
becomes important in $\Delta_D$, as well as the second order
iteration in
$(2m_c)^{-1} <\vec{\sigma} \vec{B}>$. Assuming that both effects are
of
the same
order
of magnitude, we can roughly estimate the matrix element $(2m_c^{-
1})
<\vec{\sigma} \vec{E} \times \vec{\pi}>$ as the difference between
$\Delta_D$ and $\Delta_B$,
\begin {equation}
\mid (2m_c)^{-1}<\vec{\sigma} \vec{E} \times \vec{\pi}> \mid \leq
\Delta_D - \Delta_B \approx 0.04 \ {\rm{ GeV}}^2
\label {eq:massunc}
\end {equation}
Next, observe that
\begin {equation}
\frac{i}{2} \bar{c} \sigma_{\mu\nu} G^{\mu\nu} c =
-\bar{c} \vec{\sigma} \vec{B} c
-\frac{1}{m_c} \bar{c} \vec{\sigma} \vec{E} \times \vec{\pi} c
-\frac{1}{2m_c}\bar{c} (D_iE_i) c
\label {eq:sigexp}
\end {equation}
The last term in the above equation reduces to the four fermion
operator
which we can take into account explicitly. The second term will be
estimated
as an uncertainty in the expression relating $<\frac{i}{2}\sigma G>$ to
$\Delta_D$,
$$
\mu_G^2 \equiv <\frac{i}{2} \sigma G> = \Delta_D \pm 2(\Delta_B
-\Delta_D)
-(2m_c)^{-1} 4\pi\alpha_s <\bar{c}\gamma_{\mu}
t^ac\bar{q}\gamma_{\mu}t^aq>.
$$
Using factorization for the ${\cal O}_6$ term above, and the same
values
for the parameters as above, we get  $+0.01$ for the contribution of
${\cal O}_6$ so
that
$$
\mu_G^2 = \Delta_D \pm 2(\Delta_B - \Delta_D) = 0.42 \pm 0.08 \
{\rm GeV}^2
$$
As for $\mu_{\pi}^2$, we will assume that the error bars
in Eq. (10) give the estimate of the $1/m_c$ part in
$\mu_{\pi}^2$. Moreover, it was shown previously that the sign of
the $1/m_c$
correction in $\mu_{\pi}^2$ is negative (see Eqs. (41) and (42) in Ref.
[26]).

\subsection{Factorization and estimate of the matrix element}

We
must estimate the matrix elements of the transition operator
over the  $D$-meson state.
As was mentioned above, to this end we use the factorization
procedure. We realize, of course, that it is not exact --
deviations from factorization definitely exist --
still it seems safe to say that it  gives a reasonable estimate of the
four-fermion operators, especially as far as the signs are concerned.

The relevant operators are first rearranged, through the  Fierz
transformations,  into the
form $\bar c \Gamma_\mu q\bar q\Gamma^\mu c$, with
appropriate
coefficients. Then the  matrix elements of the latter operators are
found by saturating with the vacuum intermediate state,
\begin{equation}
<D\vert \bar c \Gamma_\alpha q\bar q\Gamma^\alpha c\vert
D>=f^2_DM^2_D\, ,
\label{fact}
\end{equation}
\begin{equation}
<D\vert \bar c \Gamma_\alpha t^aq\bar q\Gamma^\alpha t^ac\vert
D>=0
\label{fact2}
\end{equation}
Here we used the definition
$$
<0\vert \bar q\Gamma_\alpha c\vert D>=if_Dp_\alpha\, .
$$
In Eq. (\ref{fact2}) we accounted for the fact that the matrix element
of the color current between the vacuum and $D$ meson  is zero.

The $L-R$ chiral  structure in Eq. (20) is crucial. We
face here  an exact
 analogy with the usual penguins. Indeed, if we use factorization
while estimating the relevant four-quark matrix elements, we see
that the contribution of the left-handed light-quark current
 is zero, so we can use Eq. (16). The corresponding result is that of
the
factorized transition operator of Eq. (20).

Let us parenthetically note that the penguins  we obtain here
  have no relation to the penguin operators of the type
 $$
(\bar ct^a s)_L\sum_f (\bar q_ft^aq_f)_{L+R}
 $$
 contributing   to charm nonleptonic decays. Although our penguins
look similar, their origin
 is completely different from the usual ones.

After summing up all new effects, i.e. the effects coming from both
the
expansion of ${\hat T}$ and the uncertainty of ${\cal O}_G$
and ${\cal O}_{\pi}$ we get, pushing
things to the high side,
\begin {equation}
\frac {\Delta \Gamma}{\Gamma} = -0.06 \pm 0.06 + 0.03
\label {result}
\end {equation}
here, the first number is due to the four-quark terms in the
transition operator, the second is
due to the uncertainty of ${\cal O}_G$, and the third is due to
${\cal O}_{\pi}$.

According to our estimates, the uncertainty in $ {\cal O}_G$ and
${\cal O}_\pi$
is enough to possibly make the total contribution of $O(1/m_c^3)$
roughly zero.
The sign of the ${\cal O}_G$ corrections is undetermined, however,
and a minus
sign gives a result which worsens the agreement with experiment.

\section{Ways out}

We saw in the previous section that contrary to all the hopes,
the $1/m^3_c$ contribution to the inclusive semileptonic width,
however exotic it is,  does not solve the problem of the deficit of the
semileptonic inclusive width.
There are several possibilities which might explain  why the
general heavy quark expansion
fails to reproduce the experimental width.

First, the factorization that we used while estimating matrix
elements can be suspected. However, the corrections to factorization
can be estimated using the method of ref. \cite{BS2} and
they seem small.

Second,  a possibility exists that operators of dimension 7
are important. In principle, this may happen since the expansion
parameter is $\sim \sqrt{\mu_\pi^2}/m_c\sim 0.7$ and is of order
unity. However, since the correction due to the dimension-6
operators is roughly only $10\%$ it seems unlikely that
the dimension-7 contribution will dominate.

At the moment it seems most probable to us that
the discrepancy demonstrated above is explained by the fact that the
family of charm lies below the duality domain.
The violations of the quark-hadron duality can be viewed as a
cumulative effect of all
high-dimension operators, taken together. Let us elucidate this
assertion in more detail.

Constructing the transition operator as an expansion in
$m_c^{-1}$  we rely on the Wilson operator product expansion.
OPE is well-formulated in the Euclidean domain
where all field fluctuations can be classified as short-distance and
large distance. Even in the Euclidean domain the divergence of
the non-perturbative series in $1/m_c$ in high orders produces
exponential terms of the type $\exp (-m_c^\gamma )$
which are not seen to any finite order in the expansion
\cite{Shifman}. To get a rough idea of these
terms one has to invoke instantons or similar
model considerations. From the QCD sum rules
it is known \cite{SVZ} that these terms are essentially unimportant
in
the Euclidean domain till surprisingly low off-shellness.

Kinematics of the problem at hand is essentially Minkowskean
since we have to take the imaginary part of the transition operator at
the very end. One justifies an OPE-based procedure by keeping in
mind an analytical continuation.
In the problem of the semileptonic width this may be a continuation
in the momentum of the lepton pair \cite{3} --
one considers the transition operator at such momenta
that one is actually off the cuts corresponding
to production of the hadronic states, in the Euclidean domain.
The prediction on the cuts is made by invoking dispersion relations,
in full analogy with what is usually done in the problem
of the total hadronic cross section in the $e^+e^-$ annihilation.
In general, one can analytically continue in some
auxiliary momenta which has nothing to do with any of
the physical momenta. This becomes the only option, say,
in the problem of inclusive nonleptonic widths.

Whatever analytic continuation is done, strictly speaking
the prediction for each given term in $1/m_c$ expansion refers
to the Euclidean domain and is translated to the Minkowski domain
only in the sense of averaging which occurs automatically
through the dispersion relations. If the integrand is smooth, however,
we can  forget about the averaging, because in this case
smearing is not needed. This is what happens, in particular,
with the total hadronic cross section in the $e^+e^-$ annihilation
at high energies -- the quark-hadron duality sets in and
the OPE-based consideration yields the value of the cross
section at a given energy, locally (without smearing).  At what
energy release is the integrand
smooth and can the terms in the $1/m_c$ expansion can be predicted
locally? The existing theory gives no answer to this question.
It may well happen that at a given (Minkowskean) energy $E$
the deviations from duality fall off only as a power of $1/E$.
Such a regime takes place, for instance, in a model discussed in ref.
\cite{alpha}. This model is definitely relevant for the large $N_c$
limit. It seems more likely, however, that in the real QCD,
 with $N_c=3$, the violations of duality fall off
exponentially,  $\exp (-(E/E_0)^\gamma )$, and the rate of this
fall off is correlated with the divergence of the high-order terms in
the power series \cite{Shifman}.

Very little is known about this aspect of QCD at present, the issue
definitely deserves further study which is clearly beyond the scope
of the present paper devoted to an applied problem. The genuinely
theoretical approach would require determining the rate of the
divergence of the high-order terms in the power expansion. Alas, we
can not do that, and the $1/m_c^3$ term found gives no hint on this
divergence whatsoever.

In the absence of theoretical considerations we are forced to rely on
phenomenological information. The total hadronic width in
$\tau$ decays is a problem close in essence to that considered here.
Moreover, the $\tau$ mass is only slightly higher than $m_c$.
A detailed QCD-based analysis of $\tau$ decays has been carried out
in ref. \cite{Pich}, and the discussion of the results of ref. \cite{Pich}
in the context the issue of duality is given is \cite{alpha}.  There are
good reasons to believe that the deviation from duality in the $\tau$
hadronic width is at the level of 7\%.  Is it then reasonable to expect
that descending from $m_\tau$ to $m_c$ we get a deviation at the
level of factor of 1.5 or 2?

The fact that the onset of duality is not universal, generally speaking,
and depends on the channel considered,
is known for a long time \cite{Novikov2}. Where specifically
the difference lies between $\tau\rightarrow\nu X$ and
$D\rightarrow l\nu X_s$ (apart from  the obvious difference,
$m_c/m_\tau\sim 0.7$) remains to be found.

Finally, let us emphasize that all attempts to determine
parameters of
QCD or HQET from the analysis of the heavy quark expansion in the
charm family   must be viewed
with
extreme caution and are hardly reliable in view of the uncontrollable
theoretical situation discussed above.

\vspace*{0.3cm}

{\bf ACKNOWLEDGMENTS} \hspace{0.4cm} We are grateful
to N. Uraltsev for useful
comments. This work was supported in part by DOE under the
grant
number DE-FG02-94ER40823 and by the Israel Science Foundation
administered
by the Israel Academy of Sciences and Humanities. Dave Dikeman is
supported
by a DOE GAANN Fellowship, and would like to thank Lev Koyrakh
for his unending
enthusiasm for discussion.


\section{Appendix}

\par In this Appendix we discuss the calculation of the
transition operator associated with Fig. 2b. We use the
Fock-Schwinger gauge
$x^\mu A_\mu^a =0$ and the background field method (see Ref.
\cite{NSVZ} for details). We shall need
the part of
 the propagator  $S(x,0)$, see Eq.(17), for
the quark  with mass $m$ in external gluon field
that contains odd number of $\gamma$ matrices
and is proportional to  $DG$.
Note that in the limit $m\rightarrow 0$ the propagator becomes
singular,
\begin{equation}
\Delta S\rightarrow {\rm ln}(m\sqrt{(-x^2)})
\label{eq:A2}
\end{equation}
The contribution of diagram of Fig. 2b to the transition operator
is equal to
\begin{equation}
\hat {T}_{\rm Fig. 2b}= {\rm Im}\vert V_{cs}\vert^2 \frac{
G^2_F}{2}\int e^{ipx}
\bar c \Gamma_\mu\Delta S(x,0)\Gamma_\nu c(0) ({\rm
tr}[S(0,x)\Gamma^\nu
S(x,0)\Gamma^\mu])
\label{eq:A3}
\end{equation}
where the trace term in the brackets represents the lepton loop, and
$\Delta S$ stands for the $DG$ part of the $s$-quark Green function.
Here $S(x,0)$ is the free massless quark propagator:
\begin{equation}
S(x,0)=\hat x/(2\pi^2x^4)
\label{eq:A4}
\end{equation}
The direct calculation uses the formulae
of Ref. \cite{BB} for the imaginary part
of the Fourier transform of the integrals of the type
\begin{equation}
\int K_n(m\sqrt{(-x^2)})/(-x^2)^p d^4xe^{ipx}.
\label{eq:A5}
\end{equation}
We just write Eq. (\ref{eq:A2}) explicitly
in $x$ space using the propagators from Eqs. (\ref{eq:A1}) and
(\ref{eq:A4})
and then use the formulae of Ref. \cite{BB} to convert
the resulting expression into the  imaginary
part of the integral in Eq. (\ref{eq:A3}).
We then see that the diagram of Fig. 2b gives the following
contribution to
Im $\hat {T}$:
\begin{equation}
\frac{-\vert
V_{cs}\vert^2G^2_F}{3^22^5\pi^3}(p^2g^{\nu\beta}+2p^{\nu} p^{\beta})
\bar c(0)\gamma_\nu (1+\gamma_5)D^\alpha G_{\alpha\beta}c(0)
({\rm ln}(m^2_c/\Lambda^2)-5/6)
\label{eq:32}
\end {equation}
We now go the reference frame
connected with the center of mass of the heavy quark
 where $p^\nu=p^\beta = m_c(1,0,0,0)$. Note also that for heavy
c-quark
\begin{equation}
\bar c\gamma_0 D_\alpha G_{\alpha 0} c =\bar c (0)\gamma_\beta
D_\alpha G^{\alpha\beta}c
\label{eq:A7}
\end{equation}
The Eq. (\ref{eq:A7}) follows from the fact that
$\bar c\gamma_i c\sim {\cal O}(1/m_c)$
We finally obtain the following
'piece' of the transition operator
\begin{eqnarray}
\begin{array}{cl}
&Im \ \hat {T}_{\rm Fig. 2b}=G^2_F\vert V_{\rm
cs}\vert^2\{\frac{\alpha_s}{72\pi^2}
m^2_c({\rm ln} (m^2_c/\Lambda^2)-5/6)\\
&\\
&2(\bar c \Gamma_\beta t^a c+\bar c\gamma_\beta t^a c)\bar
q\gamma_\beta
t^a
q\\
\end{array}
\label{eq:A8}
\end{eqnarray}
The logarithmic term comes from the singularity, Eq. (\ref{eq:A2}).
This is not all however. Part of the contribution to
the transition operator of Eq. (\ref{eq:A8}) comes from the infrared
domain, and its contribution has nothing to do with the contribution
we are interested in. The infrared contribution is given by the
contracted
loop of Fig. 4. It is easily calculated, giving contribution
\begin{eqnarray}
\begin{array}{cl}
{\rm Im} \ \hat {T}_{\rm Fig. 4}=&G^2_F\vert V_{\rm
cs}\vert^2\{\frac{\alpha_s}{72\pi^2}
m^2_c({\rm ln} (\mu^2/\Lambda^2)-3/2)\\
&\\
&(\bar c \Gamma_\beta t^a c+\bar c\gamma_\beta t^a c)\bar
q\gamma_\beta
t^a
q\\
\end{array}
\label{eq:A9}
\end{eqnarray}
Subtracting ${\rm Im} \ \hat{T}_{\rm Fig. 4}$ from ${\rm Im} \
\hat{T}_
{\rm Fig. 2b}$
we get
the transition operator
$\rm {Im} \ \hat{T}_{\rm DG}$ of Eq. (20).

\vspace*{1cm}

{\bf Figure Captions}

\vspace*{0.5cm}

{\em Fig. 1.} The four-fermion term in the transition operator
as it appears at the level $\alpha_s^0$.

\vspace*{0.2cm}

{\em Fig. 2.} $(a)$ The penguin graph for the four-fermion operator
in $\hat T$. $(b)$ The diagram with the $DG$ term in the $s$ quark
line.

\vspace*{0.5cm}

{\em Fig.3.} The diagram with the free $s$ quark line
determining ${\hat T}_{free}$.

\vspace*{0.5cm}

{\em Fig. 4.} Subtraction of
the infrared part from the coefficient of the operator
$\bar c DG c$, see Fig. 2b.

\end{document}